# Directly measuring mode purity of single component in superposed optical vortices

**CHEN WANG,**[1,2] **YUAN REN,**[1,2,*] **TONG LIU** [1,2] **ZHENGLIANG LIU,**[1,2] **SONG QIU,**[1,2] **YOU DING,**[1,2] **RUIJIAN LI,**[1,2] **AND JIE ZHAO,**[1,2]

[1] *Department of Aerospace Science and Technology, Space Engineering University, Beijing 101416, China*
[2] *Lab of Quantum Detection & Awareness, Space Engineering University, Beijing 101416, China*
*\* renyuan_823@aliyun.com*

**Abstract:** Mode purity is an important reference for the quality of the optical vortex. In this work, we propose a self-interference method to directly measure single component in superposed optical vortices based on phase-shifting technology. This method has excellent flexibility and robustness, which can be applied to a variety of occasions and harsh conditions. Careful alignment and optimized error analysis allow us to generate and measure single component with mode purity as high as 99.997%. © 2020 Optical Society of America

## 1.  INTRODUCTION

An optical vortex (OV) with the spiral wavefront is a kind of structured light which carries an orbit angular momentum (OAM) per photon of $m\hbar$, where $m$ is topological charge (TC) and $\hbar$ denotes reduce Planck constant. In 1989, Coullet *et al*. proposed a kind of optical field bearing some analogy with the superfluid[1]. In 1992, Allen *et al*. showed that photons can carry OAM[2], which has aroused widespread concern among researchers [3-18]. Since then, OVs have been utilized in a plethora of applications in the field of optical micro-manipulation[19, 20], plasmonics[21, 22], and rotation speed measurement via the optical rotation Doppler effect[23-29]. The superposed OVs are often used here instead of a single component. Before practical application, we need to measure the mode purity of the superposed OVs. Since mode purity is an important reference for the quality of the OV, which determines the performance of OV in various applications. In rotating Doppler effect[30, 31], low mode purity means frequency spectrum expansion, which greatly increases the difficulty of signal. In the field of micro-manipulation, high mode purity is more conducive to the use of superposed OVs to achieve static capture of particles[32].

There are many methods to achieve measuring mode purity of superposed OVs[33-35]. But the most versatile approach is to resolve a field into a coherent sum of modes, each with a particular amplitude weighting and phase: a so-called modal decomposition[36]. However, we need to use an SLM to scan multiple holograms one by one to complete the measurement. Is there a way to complete the scanning process with a computer to improve measurement efficiency? The answer is yes. Recent work showed that the phase-shifting holography in an interferometer can be used to measure the mode purity of OVs[37]. This work required a separate reference path that increased the complexity of the setup and introduced additional phase noise. In 2019, Andersen *et al*. modified the work mentioned above to achieve the same goal with only one path[38]. However, these two works can only measure the mode purity of single OV, and they can't do anything about superposed OVs.

In this work, we propose a self-interference method to directly measure single component in superposed optical vortices based on phase-shifting technology. Since we use part of the OVs to be measured as the reference light, though use phase-shifting technology, no additional reference beams are required for interference anymore, which has not been reported before.

## 2. THEORY AND METHODS

### A PHASE-SHIFTING THEORY OF SYMMETRY SUPERPOSED OVS

The self-interference method requires the measurement of four intensities $I(x, y; \phi_s)$ of superposed OVs. $\phi_s$ denotes the shifting-phase of one single component compared with another. For simplicity without loss of generality, we only consider the situation that the superposed OVs only have two symmetry components.

May the two components of optical vortices be

$$E_1 = A_1 \exp(i\phi)\exp(i\phi_s) \\ E_2 = A_2 \exp(-i\phi) \quad (1)$$

where $\phi_s = 0, \pi/2, \pi, 3\pi/2$. We can get four intensity distributions of superposed OVs:

$$I(x, y; 0) = I'(x, y) + I''(x, y)\cos(2\phi); \quad I(x, y; \pi/2) = I'(x, y) - I''(x, y)\sin(2\phi) \\ I(x, y; \pi) = I'(x, y) - I''(x, y)\cos(2\phi); \quad I(x, y; 3\pi/2) = I'(x, y) + I''(x, y)\sin(2\phi) \quad (2)$$

where $I'(x, y) = A_1^2 + A_2^2$, $I''(x, y) = 2A_1 A_2$. Consequently, we can calculate the phase $\phi$:

$$\phi = \frac{1}{2}\tan^{-1}\left[\frac{I(x, y; 3\pi/2) - I(x, y; \pi/2)}{I(x, y; 0) - I(x, y; \pi)}\right] \quad (3)$$

Although the intensity of the two components does not affect the final result, according to interference contrast:

$$\gamma = \frac{I_{max} - I_{min}}{I_{max} + I_{min}} \quad (4)$$

we set the two equals, in order to get the best interference pattern. Where

$$I_{max} = A_1 + A_2; \quad I_{min} = A_1 - A_2 \quad (5)$$

### B PHASE-SHIFTING THEORY OF ASYMMETRY SUPERPOSED OVS

We further promote the results. If the two components carry different OAM, that is

$$E_1' = A_0 \exp(i\phi_1)\exp(i\phi_s) \\ E_2' = A_0 \exp(i\phi_2) \quad (6)$$

where $\theta$ denotes azimuth angle. The result of phase-shifting is $\phi' = \phi_1 - \phi_2$. If the TCs of superposed OVs are $m_1$ and $m_2$, we can get the final results are

$$\phi_1 = \frac{m_1}{m_1 - m_2}; \quad \phi_2 = \frac{m_2}{m_1 - m_2} \quad (7)$$

### C GENERATION OF PHASE-SHIFTING SUPERPOSED OVS

In this part, we briefly introduce the process of generating phase-shifting superposed OVs. To obtain four intensities of phase-shifting superposed OVs, four corresponding holograms are required. Each hologram is encoded with the sum of the phase-shifting superposed OVs and the blazed grating, as shown in Fig 1. The blazed grating is to prevent part of the unmodulated light from mixing into the required superposed OVs, diffracting the required light beam to the first order, and leave the unmodulated light in the zero order. This is due to the gaps in the SLM liquid crystal arrangement. The amplitude modulation is also added to the actual operation employed with the technique shown in ref.[39, 40]. Although using an SLM for complex amplitude modulation sacrifices phase depth, it allows us to radially modulate the incident light field to generate an eigenmode of OVs instead of hypergeometric mode[41]. Load the encoded holograms onto the SLM and we can get phase-shifting superposed OVs.

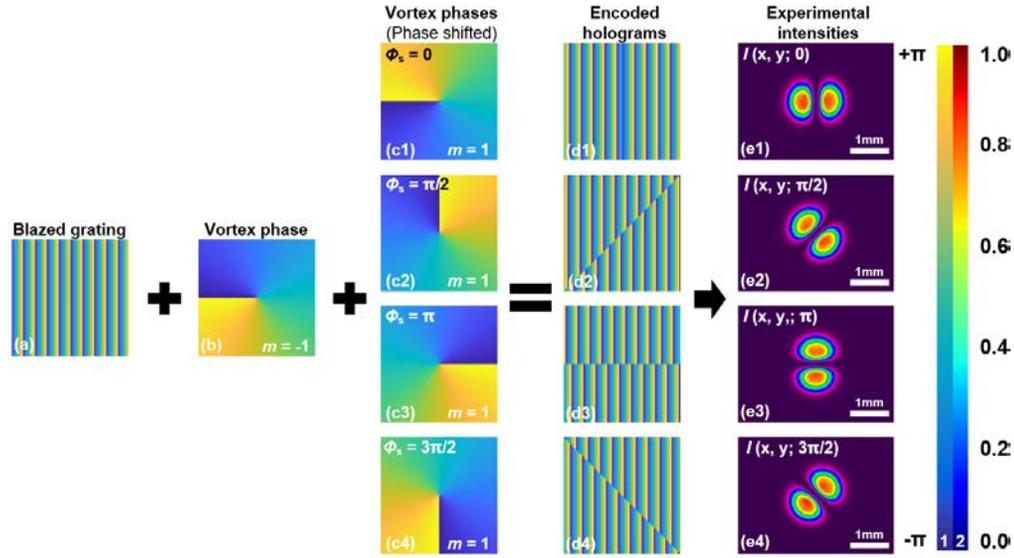

Fig 1 Schematic representation of generating phase-shifting superposed OVs. The TCs of superposed OVs here are $\pm 1$. (a) The phase of a blazed grating. (b) The phase of one component in superposed OVs. (c1-c4) Phases of another component of superposed OVs after phase-shifting. (d1-d4) Encoded holograms. The amplitude modulation is also added to the actual operation. (e1-e4) Experimental intensities of phase-shifting superposed OVs generated by (d1-d4). The color bar 1 indicates the phases range. The color bar 2 indicates the range of intensities.

## D CALCULATING THE PHASE OF SINGLE COMPONENT IN SUPERPOSED OVS

Theoretically speaking, after obtaining four intensity distribution results, the desired phase can be deduced inversely. However, in the actual calculation, due to the introduction of trigonometric functions, the desired phase is wrapped in $2\pi$. We cannot get the desired result by simply dividing by 2, as shown in Eqs.(3). Consequently, the phase unwrapping technology is required[42]. Firstly, we need to trim the phase to achieve maximum efficiency of phase unwrapping, as depicted in Fig 2(b) and Fig 2 (c). Finally, we can get the unwrapping phase by employed the phase unwrapping technology shown in ref.[42], as depicted in Fig 2(e1) and Fig 2(e2). We can directly perform phase unwrapping, or perform windowed Fourier transform[43] (WFT) first and then perform phase unwrapping. The WFT is adopted here to filter the trimmed phase to further reduce noise and smooth the phase. The unwrapped phase obtained directly in the laboratory environment is ideal, but it does not mean that good results are maintained under complex conditions such as outdoors. Therefore, The WFT is optional according to the actual situation.

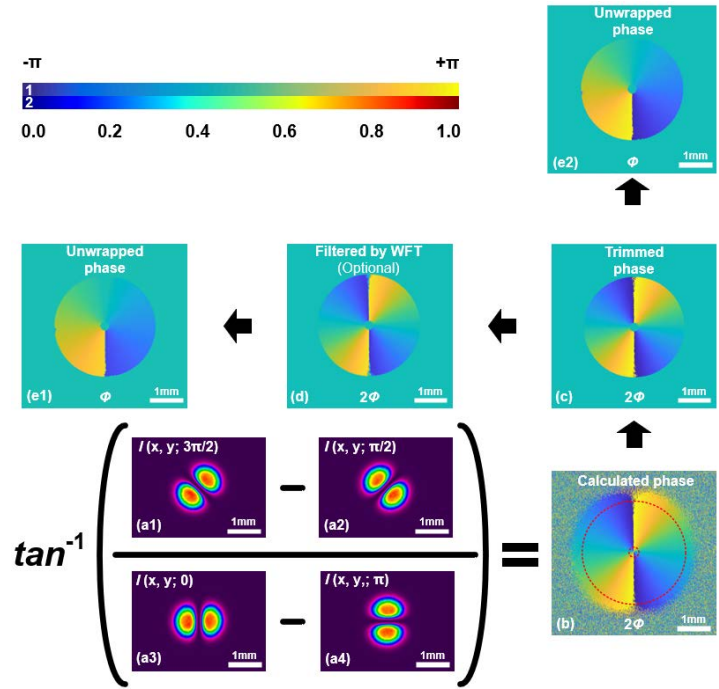

Fig 2 Schematic representation of calculating required phases. The TCs of superposed OVs here are $\pm 1$. (a1-a4) Intensities of four phase-shifting superposed OVs. (b) Calculated phase. The red dotted lines indicate the trimming area. (c) Trimmed phase. (d) Trimmed phase filtered by WFT (optional). (e1-e2) Unwrapped phase. The color bar 1 indicates the phases range. The color bar 2 indicates the range of intensities.

## 3. EXPERIMENTAL RESULTS

The experimental setup employed to generate a PPOV is shown in 错误!未找到引用源。. The laser (NEWPORT N-LHP-151) delivers a collimated Gaussian beam with wavelength of 632.8nm after a linear polarizer (LP), a half-wave plate (HWP), and a telescope consisting of two lenses (L1, L2) are used for collimation. The combination of the LP and the HWP is served to rotate the laser polarization state along the long display axis of SLM and adjust the power of incident light on SLM. The SLM (UPOLABS HDSLM80R) precisely modulates the incident light via loading a hologram mentioned above, and then the aperture (AP) is used to select the first diffraction order of the beam to avoid other stray light. A CCD camera (NEWPORT LBP2) registers the intensity pattern after L4.

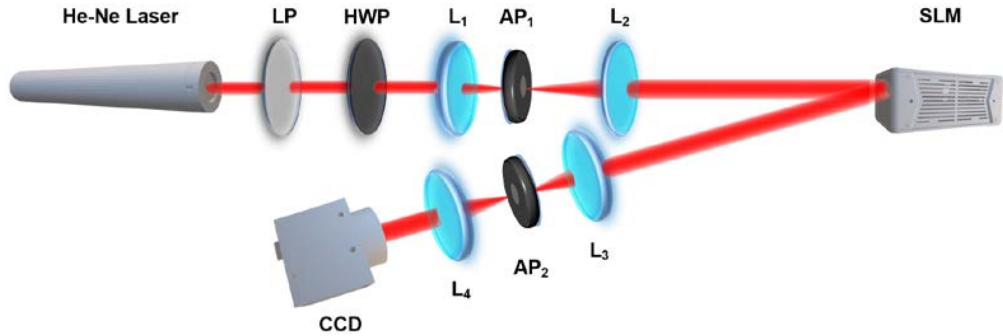

Fig 3 Experimental setup for measuring mode purity of single component in superposed. LP: linear polarizer. HWP: half-wave plate. L: lens. SLM: spatial light modulator. AP: aperture. CCD: charge-coupled device.

Experimental results of phase unwrapping are depicted in Fig 4. According to the distributions of OAM, in general, all situations can be divided into symmetrical distribution and asymmetrical distribution. These two situations can be subdivided into four categories. In the situation of symmetrical distribution, one is radial node $p$ existed, and another is high order superposed OVs. In the situation of asymmetrical distribution, one is $|m_1 - m_2| \leq |m_1|$ or $|m_2|$, and another is $|m_1 - m_2| \geq |m_1|$ and $|m_2|$. We are going to demonstrate the process of phase unwrapping in different situations below.

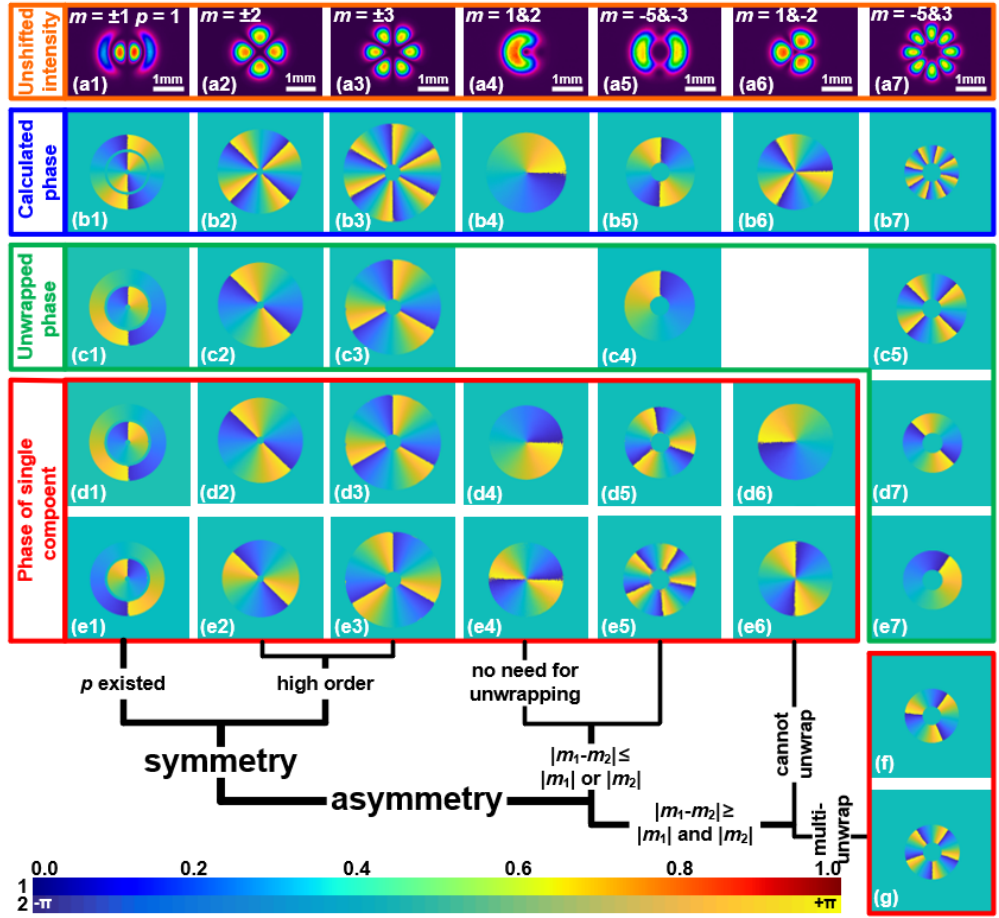

Fig 4 Experimental results of phase unwrapping. (a1-a7) Unshifted intensities. (b1-b7) Calculated phases. (c1-c5, d7, e7) Unwrapped phases. (d1-d6, e1-e6, f, g) Phases of single component. The color bar 1 indicates the range of intensities. The color bar 2 indicates the phases range. The tree diagram below the figure identifies the different situations of phase calculation.

In the situation of symmetrical distribution, as an effective complement to the example as shown in Fig 1, we demonstrate the versatility of this method when the radial node $p$ existed, as shown in Fig 4(a1-e1). At the same time, this method is also suitable for high-order superposed OVs, as depicted in Fig 4(a2-e2) and Fig 4(a3-e3). In the situation of asymmetrical distribution, when $|m_1 - m_2| \leq |m_1|$ or $|m_2|$, the phase processed by the phase-shifting technology is the required phase, and the phase unwrapping is no longer required, as shown in Fig 4(a4-e4). Correspondingly, when $|m_1 - m_2| \geq |m_1|$ and $|m_2|$, since the calculated phase is the OV of odd-numbered TCs, the phase cannot be unwrapped, as shown in Fig 4(a6-e6). So, it is necessary to directly subtract the phase from the calculated phase. This operation retains the error of phase-shifting technique for phase calculation, Consequently, it does not affect the result theoretically. Finally, there is a special situation that needs to be introduced: multiple phase unwrapping is required in the process of calculating the phase, as shown in Fig 4(a7-e7, f, g). We have performed phase unwrapping three times in total to get the phase distribution of the fundamental mode. To get the final phase distribution, we also need to multiply the corresponding TCs $-5$ and $3$.

We selected 4 groups of representative phases to calculate the mode purity. The mode purity for an arbitrary field $\Psi$ can be calculated with[44, 45]:

$$\Gamma_p = \frac{\sum_{m=-\infty}^{+\infty} C_m^p}{\sum_{p=0}^{+\infty}\sum_{m=-\infty}^{+\infty} C_m^p}, \Gamma_m = \frac{\sum_{p=0}^{+\infty} C_m^p}{\sum_{p=0}^{+\infty}\sum_{m=-\infty}^{+\infty} C_m^p} \quad (8)$$

where

$$C_m^p = \int_0^\infty \left\langle a_m^p(r,\phi,z)^* | a_m^p(r,\phi,z) \right\rangle r dr \quad (9)$$

$$a_m^p(r,z) = 1/(2\pi)^{1/2} \int_0^{2\pi} \Psi(r,\phi,z) \mathrm{LG}_p^m d\phi \quad (10)$$

The calculated mode purity of single component is shown in Fig 5. Whether it is radial decomposition or angular decomposition, the model purity can be calculated by self-interference method. Experimental results show that the purity of the angular mode (OAM spectrums) is one order of magnitude higher than that of the radial mode (radial distributions), which is consistent with the more robustness of the angular distribution of an OV.

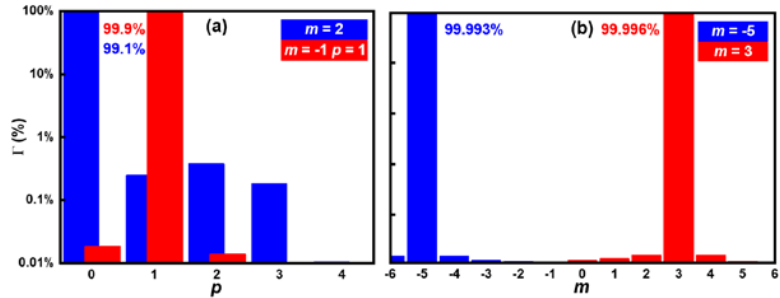

Fig 5 Experimental distributions of single component mode purity. (a) Radial distributions of $m = 2$ ($p = 0$) and $m = -1$ ($p = 1$) from superposed OVs $m = 1 \& 2$ ($p = 0$) and $m = \pm 1$ ($p = 1$), respectively. OAM spectrums of OVs $m = -5$ and $m = 3$ from superposed OVs $m = -5 \& 3$ and $m = \pm 3$, respectively.

## 4. DISCUSSION

In this part, we would like to discuss the effect of experimental error on the results. Firstly, we demonstrate the effect of different exposure times on the measurement. The results show that as the exposure time increases, the measured mode purity decreases slightly, as shown in Fig 6(a). This can be explained that without affecting the sampling, the increase in exposure time causes the noise in the image to increase, so the mode purity is slightly reduced. Secondly, we demonstrate the effect of CCD tilt on the measurement. The results show that as the CCD tilt increases, the measured mode purity decreases slightly, as shown in Fig 6(a). This can be explained that the tilt of CCD affects the symmetry distribution of superposed OVs, which in turn affects the accuracy of the phase calculation, as shown in Fig 6(b). Finally, we demonstrate the effect of CCD shift on the measurement. The results show that as the CCD shift increases, the measured mode purity decreases slightly, as shown in Fig 6(c). Although the CCD shift does not change the distribution of the superposed OVs, if we still perform mode decomposition at the position shown in 0 pixel, the accuracy of the mode decomposition will be reduced.

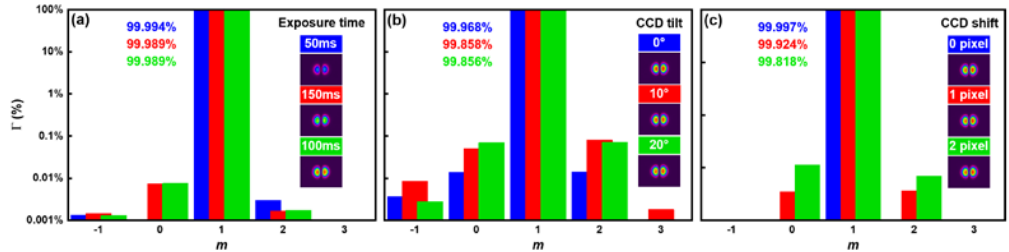

Fig 6 The effect of experimental error on the results. The effect of (a) exposure times, (b) CCD tilt and (c) CCD shift on the measurement of the mode purity.

**Funding**

This work was supported in part by the National Nature Science Foundation of China under Grant 11772001 and 61805283, and in part by the Key research projects of the foundation strengthening program and Outstanding Youth Science Foundation.

**Disclosures**

The authors declare no conflicts of interest.

**References**

1. P. Coullet, L. Gil, and F. Rocca, "Optical vortices," Optics Communications **73**, 403-408 (1989).
2. L. Allen, M. W. Beijersbergen, R. J. Spreeuw, and J. P. Woerdman, "Orbital angular momentum of light and the transformation of Laguerre-Gaussian laser modes," Physical Review A Atomic Molecular & Optical Physics **45**, 8185 (1992).
3. L. Chen, W. Zhang, Q. Lu, and X. Lin, "Making and identifying optical superpositions of high orbital angular momenta," Physical Review A **88**(2013).
4. X. Li, Y. Tai, F. Lv, and Z. Nie, "Measuring the fractional topological charge of LG beams by using interference intensity analysis," Optics Communications **334**, 235-239 (2015).
5. S. Fu, C. Gao, Y. Shi, K. Dai, L. Zhong, and S. Zhang, "Generating polarization vortices by using helical beams and a Twyman Green interferometer," Opt Lett **40**, 1775-1778 (2015).
6. X. Ke and G. Tian, "Experiment on Generation of Vortex Light with Few-Mode Fiber," Chinese Journal of Lasers **44**(2017).
7. H. Ma, X. Li, Y. Tai, H. Li, J. Wang, M. Tang, Y. Wang, J. Tang, and Z. Nie, "In situ measurement of the topological charge of a perfect vortex using the phase shift method," Opt Lett **42**, 135-138 (2017).
8. X. Qiu, F. Li, W. Zhang, Z. Zhu, and L. Chen, "Spiral phase contrast imaging in nonlinear optics: seeing phase objects using invisible illumination," Optica **5**, 208 (2018).
9. D. Yang, Y. Li, D. Deng, Q. Chen, Y. Zhang, Y. Liu, J. Gao, and M. Sun, "Chiral optical field generated by an annular subzone vortex phase plate," Opt Lett **43**, 4594-4597 (2018).
10. Y. Yang, Q. Zhao, L. Liu, Y. Liu, C. Rosales-Guzmán, and C.-w. Qiu, "Manipulation of Orbital-Angular-Momentum Spectrum Using Pinhole Plates," Physical Review Applied **12**(2019).
11. Y. Shen, X. Wang, Z. Xie, C. Min, X. Fu, Q. Liu, M. Gong, and X. Yuan, "Optical vortices 30 years on: OAM manipulation from topological charge to multiple singularities," Light: Science & Applications **8**, 90 (2019).

12. D. Yang, Y. Li, D. Deng, J. Ye, Y. Liu, and J. Lin, "Controllable rotation of multiplexing elliptic optical vortices," Journal of Physics D: Applied Physics **52**(2019).
13. X. Fang, H. Ren, and M. Gu, "Orbital angular momentum holography for high-security encryption," Nature Photonics (2019).
14. H. Zhang, X. Li, H. Ma, M. Tang, H. Li, J. Tang, and Y. Cai, "Grafted optical vortex with controllable orbital angular momentum distribution," Optics express **27**(2019).
15. H. Ma, X. Li, H. Zhang, J. Tang, H. Li, M. Tang, J. Wang, and Y. Cai, "Optical vortex shaping via a phase jump factor," Opt Lett **44**, 1379-1382 (2019).
16. Z. Man, Z. Xi, X. Yuan, R. E. Burge, and H. P. Urbach, "Dual Coaxial Longitudinal Polarization Vortex Structures," Physical Review Letters **124**, 103901 (2020).
17. Y. Zhai, S. Fu, J. Zhang, Y. Lv, H. Zhou, and C. Gao, "Remote detection of a rotator based on rotational Doppler effect," Applied Physics Express **13**(2020).
18. Z. Man, P. Meng, and S. Fu, "Creation of complex nano-interferometric field structures," Optics Letters **45**, 37-40 (2020).
19. W. H. Campos, J. M. Fonseca, J. B. S. Mendes, M. S. Rocha, and W. A. Moura-Melo, "How light absorption modifies the radiative force on a microparticle in optical tweezers," Applied Optics **57**, 7216-7224 (2018).
20. H. Wang, L. Tang, J. Ma, X. Zheng, D. Song, Y. Hu, Y. Li, and Z. Chen, "Synthetic optical vortex beams from the analogous trajectory change of an artificial satellite," Photonics Research **7**(2019).
21. H. Kim, J. Park, S. W. Cho, S. Y. Lee, M. Kang, and B. Lee, "Synthesis and Dynamic Switching of Surface Plasmon Vortices with Plasmonic Vortex Lens," Nano Letters **10**, 529-536 (2010).
22. Nir, Shitrit, Itay, Bretner, Yuri, Gorodetski, Vladimir, Kleiner, Erez, and Hasman, "Optical Spin Hall Effects in Plasmonic Chains," Nano Letters **11**, 2038–2042 (2011).
23. L. Marrucci, "Spinning the Doppler effect," Science **341**, 464-465 (2013).
24. M. Padgett, "A new twist on the Doppler shift," Physics Today **67**, 58-59 (2014).
25. H. Zhou, D. Fu, J. Dong, P. Zhang, and X. Zhang, "Theoretical analysis and experimental verification on optical rotational Doppler effect," Optics express **24**, 10050-10056 (2016).
26. Hai-Long, Zhou, Dong-Zhi, Fu, Jian-Ji, Dong, Pei, Zhang, Dong-Xu, and Chen, "Orbital angular momentum complex spectrum analyzer for vortex light based on the rotational Doppler effect," Light Science & Applications (2016).
27. H. L. Zhou, D. Z. Fu, J. J. Dong, P. Zhang, D. X. Chen, X. L. Cai, F. L. Li, and X. L. Zhang, "Orbital angular momentum complex spectrum analyzer for vortex light based on the rotational Doppler effect," Light Sci Appl **6**, 16251 (2017).
28. S. Fu, T. Wang, Z. Zhang, Y. Zhai, and C. Gao, "Non-diffractive Bessel-Gauss beams for the detection of rotating object free of obstructions," Optics express **25**, 20098-20108 (2017).
29. W. Zhang, D. Zhang, X. Qiu, and L. Chen, "Quantum remote sensing of the angular rotation of structured objects," Physical Review A **100**, 043832 (2019).
30. S. Qiu, T. Liu, Z. Li, C. Wang, Y. Ren, Q. Shao, and C. Xing, "Influence of lateral misalignment on the optical rotational Doppler effect," Appl Opt **58**, 2650-2655 (2019).
31. S. Qiu, T. Liu, Y. Ren, Z. Li, C. Wang, and Q. Shao, "Detection of spinning objects at oblique light incidence using the optical rotational Doppler effect," Optics express **27**, 24781-24792 (2019).
32. V. G. Shvedov, A. S. Desyatnikov, A. V. Rode, W. Z. Krolikowski, and Y. S. Kivshar, "Optical guiding of absorbing nanoclusters in air," Optics express **17**, 5743-5757 (2009).
33. J. M. Hickmann, E. Fonseca, W. C. Soares, and S. Chávez-Cerda, "Unveiling a Truncated Optical Lattice Associated with a Triangular Aperture Using Light's Orbital Angular Momentum," Physical Review Letters **105**, 053904 (2010).
34. Zhimin, Shi, Mohammad, Mirhosseini, Jessica, Margiewicz, Mehul, Malik, Freida, and Rivera, "Scan-free direct measurement of an extremely high-dimensional photonic state," Optica **2**, 388-392 (2015).

35. G. Kulkarni, R. Sahu, O. S. Maga?A-Loaiza, R. W. Boyd, and A. K. Jha, "Single-shot measurement of the orbital-angular-momentum spectrum of light," Nature Communications **8**, 1054 (2017).
36. J. Pinnell, I. Nape, B. Sephton, M. A. Cox, V. Rodriguez-Fajardo, and A. Forbes, "Modal analysis of structured light with spatial light modulators: a practical tutorial," J Opt Soc Am A Opt Image Sci Vis **37**, C146-C160 (2020).
37. A. D'Errico, R. D'Amelio, B. Piccirillo, F. Cardano, and L. Marrucci, "Measuring the complex orbital angular momentum spectrum and spatial mode decomposition of structured light beams," Optica **4**, 1350 (2017).
38. J. M. Andersen, S. N. Alperin, A. A. Voitiv, W. G. Holtzmann, J. T. Gopinath, and M. E. Siemens, "Characterizing vortex beams from a spatial light modulator with collinear phase-shifting holography," Applied Optics **58**, 404 (2019).
39. T. W. Clark, R. F. Offer, S. Franke-Arnold, A. S. Arnold, and N. Radwell, "Comparison of beam generation techniques using a phase only spatial light modulator," Optics express **24**, 6249-6264 (2016).
40. C. Rosales-Guzmán and A. Forbes, "How to Shape Light with Spatial Light Modulators," (2017).
41. D. Alessio, D. Raffaele, P. Bruno, C. Filippo, and M. Lorenzo, "Measuring the complex orbital angular momentum spectrum and spatial mode decomposition of structured light beams," Optica **4**, 1350- (2017).
42. M. Herráez, D. R. Burton, M. J. Lalor, and M. A. Gdeisat, "Fast two-dimensional phase-unwrapping algorithm based on sorting by reliability following a noncontinuous path," Applied Optics **41**, 7437-7444 (2002).
43. K. Qian, "Two-dimensional windowed Fourier transform for fringe pattern analysis: Principles, applications and implementations," Optics & Lasers in Engineering **45**, 304-317 (2007).
44. H. I. Sztul and R. R. Alfano, "The Poynting vector and angular momentum of Airy beams," Optics express **16**, 9411-9416 (2008).
45. L. Torner, J. Torres, and S. Carrasco, "Digital spiral imaging," Optics express **13**, 873-881 (2005).